\documentclass[twoside,twocolumn,english,aps,prd,showpacs,showkeys]{revtex4}
\usepackage[T1]{fontenc}
\usepackage[latin1]{inputenc}
\usepackage{amsmath}
\usepackage{graphicx}
\usepackage{amssymb}

\makeatletter

\newcommand{\lyxdot}{.}

\usepackage{bm}

\usepackage{babel}
\makeatother
\begin{document}

\title{Searching for New Physics in Future Neutrino Factory Experiments}

\author{J.~Holeczek}

\author{J.~Kisiel}

\author{J.~Syska}

\author{M.~Zra\l{}ek}

\email{Marek.Zralek@us.edu.pl}

\affiliation{Institute of Physics, University of Silesia, ul. Uniwersytecka 4,
40-007 Katowice, Poland}

\date{\today}

\begin{abstract}
An extension of the New Standard Model, by introducing a mixing of
the low mass {}``active'' neutrinos with heavy ones, or by any model
with lepton flavor violation, is considered. This leads to non-orthogonal
neutrino production and detection states and to modifications of neutrino
oscillations in both, vacuum and matter. The possibility of the discovery
of such effects in current and future neutrino oscillation experiments
is discussed. First order approximation formulas for the flavor transition
probabilities in constant density matter, for all experimentally available
channels, are given. Numerical calculations of flavor transition probabilities
for two sets of New Physics parameters describing a single {}``effective''
heavy neutrino state, both satisfying present experimental constraints,
have been performed. Two energy ranges and several baselines, assuming
both the current ($\pm2\sigma$) and the expected in future ($\pm3\%$)
errors of the neutrino oscillation parameters are considered, keeping
their present central values. It appears that the biggest potential
of the discovery of the possible presence of any New Physics is pronounced
in oscillation channels in which $\nu_{e}$, $\nu_{\bar{e}}$ are
not involved at all, especially for two baselines, $L=3000\, km$
and $L=7500\, km$, which for other reasons are also called {}``magic''
for future $Neutrino\, Factory$ experiments. 
\end{abstract}



\maketitle

\section{Introduction}

\label{sec:intro} For several years, neutrinos are considered to
be massive particles \cite{mass}, and therefore the orthodox Standard
Model (SM) with massless neutrinos must be extended. There exist two
possibilities. Firstly, the extension of the SM can appear only at
a very high energy scale, the GUT/Planck scale, and the non-zero neutrino
masses are just the visible indication of such {}``high energy''
physics in our {}``low energy'' world. Such scenario is usually
called the New Standard Model ($\nu SM$). Secondly, some New Physics
(NP) can already be present at the $TeV$ scale, that means at energies
close to our present-day experimental facilities. The second of these
possibilities is more appealing from both, the experimental, and the
theoretical points of view. Such a {}``low energy'' NP can participate
in neutrino flavor transitions, and so it could possibly be measured
in future neutrino oscillation experiments. Then everything is in
the hands of the precision of the planned experiments. The bounds
on the NP parameters, which arise from todays experiments, are too
restrictive to give any good chance to see any effects in the present
neutrino flavor transition data, taking into account the fact that
the present precision in determination of the neutrino oscillation
parameters (of about $10\%$ \cite{tests}) effectively screens off
any possible presence of NP. However, the combined expected results
from future neutrino facilities, like $Beta\, Beam$, $Super\, Beam$,
and $Neutrino\, Factory$, should bring the neutrino oscillation parameter
errors down to about $1$--$3\%$ \cite{future_precision} and therefore
give a chance for a discovery of effects, which possibly could not
be explained by the {}``present physics''.

The potential of the NP discovery is considered in this paper. There
are many ways in which NP can modify neutrino oscillations. The non-standard
effects can directly change oscillation probabilities (the so called
{}``damping signatures'' \cite{damping}), or they can modify the
oscillation amplitudes (by non-standard Hamiltonian effects \cite{sterile}--\cite{NQD}),
where both, oscillations in vacuum, and in matter can be affected.
These possibilities have extensively been examined in the existing
literature. Thus, there are models with sterile \cite{sterile} or/and
heavy \cite{heavy} neutrinos, general models with lepton flavor violation
(LFV) \cite{LFV}, non-standard interactions \cite{NSI}, flavor changing
neutral currents \cite{FCNC}, general fermion interactions \cite{GFI}
and mass varying neutrinos \cite{MVN}. Next, there are models with
non-unitary leptonic mixing \cite{CP-violation}, violation of the
Lorentz symmetry \cite{VLS}, violation of the principle of General
Relativity \cite{PGRV} and violation of the CPT symmetry \cite{CPT}.
Finally, there are models which modify neutrino oscillations, i.e.~models
with neutrino wave packet decoherence \cite{WPD}, neutrinos' decays
\cite{ND} and neutrino quantum decoherence \cite{NQD}.

In this paper, we discuss one class of NP only, which can be obtained
by mixing of the low mass {}``active'' neutrinos with heavy ones
\cite{heavy}, or by any model with LFV \cite{LFV}, in both of which
neutrinos interact with matter particles by the left-handed (L-H)
charge and neutral currents only. In such models, the effective mixing
matrix is non-unitary, resulting in non-orthogonal neutrino production
and detection states. This non-orthogonality by itself modifies neutrino
oscillations in vacuum. Apart from this, the neutrino interactions
with matter particles are non-standard and so the oscillation effects
in matter are further modified as well. Both these effects are here
taken into account. Additionally, in our numerical calculations, we
assume both $\pm3\%$ and $\pm2\sigma$ errors of today's $\nu SM$
neutrino oscillation parameters, keeping their present central values
(see Chapter~\ref{sec:sims}). The flavor transition probabilities
were calculated for two energy ranges and several baselines \cite{baselines}.
Two of these baselines, $L=3000\, km$ and $L=7500\, km$, are called
{}``magic'' for future $Neutrino\, Factory$ experiments, as they
are especially useful for CP violation discovery ($L=3000\, km$)
\cite{Superbeams} or optimal for resolving the oscillation parameters
degeneracy problem ($L=7500\, km$). We have performed our numerical
calculations of flavor oscillation probabilities for all available
channels for two sets of NP parameters which describe a single {}``effective''
heavy neutrino state, both satisfying present experimental constraints.
One of the easiest channels, from the experimental point of view \cite{first-ch},
is the $\nu_{\mu}\rightarrow\nu_{e}$ one, but it will be difficult
to observe any NP there (as it will also be in all another channels
in which $\nu_{e}$ or $\nu_{\bar{e}}$ are involved). In two other
channels, $\nu_{\mu}\rightarrow\nu_{\tau}$ and $\nu_{\mu}\rightarrow\nu_{\mu}$
(and in the corresponding antineutrino channels), the effects of NP
are seen with the largest intensity, especially for the {}``magic''
baselines mentioned above \cite{Aguila-Zralek}. In the next Chapter
we investigate how the NP modifies $\nu SM$ oscillation transition
probabilities. We discuss all leading terms, which give the new non-orthogonal
production and detection states and which modify the neutrino coherent
scattering on matter particles. We discuss also the reason why all
channels with $\nu_{\mu}$ or $\nu_{\tau}$ as the initial and final
neutrinos (and the corresponding antineutrino channels), i.e.~all
channels in which $\nu_{e}$, $\nu_{\bar{e}}$ are not involved at
all, are the most desired ones. Then, in Chapter~\ref{sec:sims},
we present the results of our numerical simulations. And finally,
in Chapter~\ref{sec:fruits}, we give our conclusions. In the Appendix~\ref{app:approx}
we collect all formulas for the flavor transition probabilities in
constant density matter for all experimentally available channels.

\section{Searching for New Physics in Neutrino Oscillation Experiments}

\label{sec:theory}If one takes into account that only relativistic
neutrinos are detected (and that only vector left-handed interactions
are considered), the detection rate $N_{\beta\alpha}$ of the $\nu_{\beta}$
neutrinos, coming from the produced $\nu_{\alpha}$ neutrinos, factorizes
into three parts, the production flux $N_{\alpha}$, the transition
probability $P_{\alpha\rightarrow\beta}$, and the detection cross
section $\sigma_{\beta}$:\begin{equation}
N_{\beta\alpha}=\, N_{\alpha}\, P_{\alpha\rightarrow\beta}\,\sigma_{\beta}\,.\label{E20}\end{equation}
 Any physics beyond the $\nu SM$ will modify all of these three parts
(or even the above factorization will be made impossible). In this
paper, we only discuss the modifications of the probability of neutrino
flavor transition from a production state $\mid\nu_{P}\rangle$ to
a detection state $|\nu_{D}\rangle$, leaving the modifications of
the production and detection neutrino cross sections for future detailed
considerations.

In the frame of the $\nu SM$, the production or detection neutrino
states are equal to the appropriate orthonormal neutrino flavor states
$|\nu_{\alpha}\rangle$. The NP modifies this dependence and therefore
the $|\nu_{P,D}\rangle$ states are only approximately equal to the
$|\nu_{\alpha}\rangle$ states. Let us assume that the neutrinos are
produced in the following process: \begin{equation}
\ell+X\rightarrow\nu+Y\,,\label{E21}\end{equation}
 where $\ell$ is a charged lepton ($\ell=e,\mu,\tau$) and $X$,$Y$
are hadrons. Then the normalized neutrino production state $|\nu_{P}\rangle$
can be defined as: \begin{equation}
|\nu_{P}\rangle\,=\frac{\sum_{i=1}^{n}A(\ell+X\rightarrow\nu_{i}+Y)\,|\nu_{i}\rangle}{\sqrt{\sum_{i=1}^{n}|A(\ell+X\rightarrow\nu_{i}+Y)|^{2}}}\,,\label{E22}\end{equation}
 where $A(\ell+X\rightarrow\nu_{i}+Y)$ is the amplitude for the process
Eq.(\ref{E21}), in which the neutrino eigenmass state $|\nu_{i}\rangle$
is produced. The sum in Eq.(\ref{E22}) goes over all neutrinos with
masses $m_{i}$ which are kinematically allowed. If particle spins
were taken into account, instead of pure states of Eq.(\ref{E22}),
we would have to use mixed states described by an appropriate density
matrix.

Let us consider a NP model, in which besides three light $\nu SM$
neutrinos there are also heavier ones, which couple to the light charged
leptons in a non-negligible way \cite{heavy}. To be more precise,
we assume the charge current Lagrangian in the following form ($n>3$):
\begin{eqnarray}
\mathcal{L}_{CC} & = & \frac{e}{2\sqrt{2}\sin\theta_{W}}\sum_{\alpha=e,\mu,\tau}\,\sum_{i=1}^{n}\label{E23}\\
 &  & \psi_{\alpha}\,\gamma^{\mu}\,(1-\gamma_{5})\,(\mathbf{U}_{\nu})_{\alpha i}\,\nu_{i}\, W_{\mu}^{-}+h.c.\,,\nonumber \end{eqnarray}
 and similarly the neutral current Lagrangian in the following form:
\begin{equation}
\mathcal{L}_{NC}=\frac{e}{2\sin(2\theta_{W})}\sum_{i,j}\overline{\nu_{i}}\,\gamma^{\mu}\,(1-\gamma_{5})\,\Omega_{ij}\,\nu_{j}\, Z_{\mu}\,,\label{E24}\end{equation}
 where \begin{equation}
\Omega_{ij}=\sum_{\alpha=e,\mu,\tau}(\mathbf{U}_{\nu})_{\alpha i}^{*}\,(\mathbf{U}_{\nu})_{\alpha j}\,.\label{E25}\end{equation}
 The $n\times n$ matrix $\mathbf{U}_{\nu}$ defines the mixing between
the flavor and mass states. So, e.g. assuming three light and three
heavy neutrinos we have: \begin{equation}
\left(\begin{array}{c}
\nu_{e}\\
\nu_{\mu}\\
\nu_{\tau}\\
N_{e}\\
N_{\mu}\\
N_{\tau}\end{array}\right)=\left(\begin{array}{cc}
\mathcal{U}_{\alpha i} & V_{\alpha I}\\
V_{ai}^{'} & \mathcal{U}_{aI}^{'}\end{array}\right)\left(\begin{array}{c}
\nu_{1}\\
\nu_{2}\\
\nu_{3}\\
N_{1}\\
N_{2}\\
N_{3}\end{array}\right)\,,\label{E26}\end{equation}
 where all submatrices ($\mathcal{U}$,$V$,$V^{'}$ and $\mathcal{U}^{'}$)
have dimensions $3\times3$. The three additional neutrino states,
$N_{e}$,$N_{\mu}$ and $N_{\tau}$, do not couple to charged leptons.
For Majorana neutrinos the submatrices $\mathcal{U}$ and $V$, which
explicitly enter the interaction Lagrangian Eq.(\ref{E23}), depend
on $3n-6$ moduli and $3n-6$ CP violating phases. For Dirac neutrinos
$n-1$ phases can be eliminated, giving altogether $2n-5$ phases
which, in principle, can enter the NP neutrino flavor transition probabilities.
If we assume that in the process given by Eq.(\ref{E21}) the energy
conservation does not allow to produce heavy neutrinos $N_{i}$ then,
according to Eq.(\ref{E22}), the neutrino production state ($|\nu_{P}\rangle$)
is given by: \begin{equation}
|\nu_{P}\rangle=\frac{1}{\sqrt{\sum_{i=1}^{3}|\mathcal{U}_{\ell\, i}|^{2}}}\sum_{i=1}^{3}\mathcal{U}_{\ell\, i}^{*}\,|\nu_{i}\rangle\,.\label{E27}\end{equation}
 Such states are normalized but not orthogonal. As the mixing of heavy
neutrinos is small ($|V_{\alpha I}|^{2}\ll1$), the matrix $\mathcal{U}$
is almost unitary. If we assume that a matrix $U$ describes a unitary
transition and is parameterized by the standard $3$ mixing angles
$\theta_{12},\theta_{13},\theta_{23}$ and $1$ standard Dirac CP
breaking phase $\delta_{13}$, then the orthonormal neutrino flavor
state $|\nu_{\alpha}\rangle$ is the following combination of neutrino
mass states $|\nu_{i}\rangle$ \begin{equation}
|\nu_{\alpha}\rangle\,=\sum_{i=1}^{3}U_{\alpha i}^{*}\,|\nu_{i}\rangle\,.\label{E28}\end{equation}
 The $\mathcal{U}$ matrix can be parameterized by a matrix $\Lambda$
close to the unit matrix $\mathbf{1}$: \begin{equation}
\mathcal{U}=\Lambda\, U\,\,\text{with}\,\,\Lambda=\mathbf{1}-\delta\Lambda\,,\label{E29}\end{equation}
 and therefore the production state $|\nu_{P}\rangle$ in Eq.(\ref{E22})
is close to the eigenflavor state $|\nu_{\alpha}\rangle$ and can
also be decomposed in the orthonormal flavor basis: \begin{equation}
|\nu_{P}\rangle\,\equiv\,|\widetilde{\nu_{\ell}}\rangle=\sum_{\alpha=e,\mu,\tau}d_{\ell\alpha}^{*}\,|\nu_{\alpha}\rangle\,,\label{E210}\end{equation}
 where the $d_{\ell\alpha}$ parameters are equal to: \begin{equation}
d_{\ell\alpha}=\frac{\Lambda_{\ell\alpha}}{\sqrt{\sum_{i=1}^{3}|\mathcal{U}_{\ell\, i}|^{2}}}\,.\label{E211}\end{equation}
 In general case, the values of parameters $(\mathbf{1}-\Lambda)_{\ell\alpha}=(\delta\Lambda)_{\ell\alpha}\equiv\delta\lambda_{\ell\alpha}$
depend on the production (detection) process \cite{LFV,BGP}, and
are bounded by the existing charged lepton data. The same parametrization
as in Eq.(\ref{E210}) was considered in \cite{LFV}, where the general
lepton flavor violation NP model is discussed. In each row of the
$d_{\ell\alpha}$ matrix, practically only one element has a non-negligible
value, namely: \begin{equation}
|d_{\ell\ell}|\leq1\,,\,\text{and}\,\,|d_{\ell\alpha}|\approx0\,\,\text{for}\,\,\alpha\neq\ell\,.\label{E212}\end{equation}
 In general however, the $3\times3$ matrix $\delta\Lambda$ can have
all non-vanishing elements. Therefore, $9$ moduli and $9$ phases
can generally parameterize any kind of NP. Not all phases play role
in the transition probabilities. Five Majorana type phases do not
enter any transition probability formula, hence only $4$ phases remain.

In the model considered here, the elements of the $\delta\Lambda$
matrix are connected with the heavy neutrino mixing matrix $V$. From
the unitary condition for the full $\mathbf{U}_{\nu}$ matrix we get
the following relation between $\Lambda$ and $V$ matrices: \begin{equation}
\Lambda\Lambda^{\dagger}=\mathbf{1}-VV^{\dagger}\,,\label{E214}\end{equation}
 so, neglecting the $\delta\Lambda\,\delta\Lambda^{\dagger}\,$ term,
we have: \begin{equation}
\delta\Lambda+\delta\Lambda^{\dagger}=VV^{\dagger}\,,\label{E215}\end{equation}
 or explicitly: \begin{equation}
\delta\lambda_{\alpha\beta}+\delta\lambda_{\beta\alpha}^{*}=c_{\alpha\beta}\,,\,\text{where}\,\, c_{\alpha\beta}=(VV^{\dagger})_{\alpha\beta}\,.\label{E216}\end{equation}

Note that, these are not explicit formulas for individual $\delta\lambda_{\alpha\beta}$
elements. However, terms in form of the left hand sides of Eqs.(\ref{E215}),(\ref{E216})
may often entirely describe the NP effects for modified matter oscillations
(see Eqs.(\ref{E218}),(\ref{E3022}) below). If we need the knowledge
of individual $\delta\lambda_{\alpha\beta}$ elements, for example
in order to calculate the new production and detection neutrino state
modifications, then they need to be calculated from Eq.(\ref{E29})
(see Eq.(\ref{E222}) below and the Appendix~\ref{app:approx}).

Up to now we have only considered the modifications of the initial
and final neutrino states in the probability formula. Such modifications
will change the neutrino propagation even in vacuum. Yet, if neutrinos
pass matter additional effects arise. The coherent neutrino scattering
on matter particles is modified by NP and because of this: (i)~neutrinos
acquire different effective masses and (ii)~their coherent scattering
amplitude is modified. These effects can be parameterized by a NP
effective Hamiltonian $\mathbf{H}^{NP}$ (see \cite{G-M}, \cite{Pramana}).
The matrix representation of $\mathbf{H}^{NP}$ operator depends on
the basis of states. Generally, in the basis of states of produced
and detected neutrinos, the $\mathbf{H}^{NP}$ operator is not represented
by a hermitian matrix. However, it is represented by a hermitian matrix
in the eigenmass basis ($|\nu_{i}\rangle$) and in any basis which
is unitary transformed, e.g. in the basis of eigenflavor orthonormal
neutrino states given by Eq.(\ref{E28}). Therefore, in the orthonormal
flavor basis ($|\nu_{\alpha}\rangle$), we can write: \begin{equation}
\mathbf{H}^{NP}=\frac{A_{e}}{2E_{\nu}}\left(\begin{array}{ccc}
\varepsilon_{ee} & \varepsilon_{e\mu}e^{i\chi_{e\mu}} & \varepsilon_{e\tau}e^{i\chi_{e\tau}}\\
\varepsilon_{e\mu}e^{-i\chi_{e\mu}} & \varepsilon_{\mu\mu} & \varepsilon_{\mu\tau}e^{i\chi_{\mu\tau}}\\
\varepsilon_{e\tau}e^{-i\chi_{e\tau}} & \varepsilon_{\mu\tau}e^{-i\chi_{\mu\tau}} & \varepsilon_{\tau\tau}\end{array}\right)\,,\label{E217}\end{equation}
 where $A_{e}=2\sqrt{2}G_{F}N_{e}E_{\nu}$ is the usual neutrino effective
amplitude, which depends on the electron matter density $N_{e}$,
and $\varepsilon_{\alpha\beta}$ and $\chi_{\alpha\beta}$ are NP
parameters, moduli and phases, which describe the effective NP neutrino
interaction with matter particles $e$, $p$ and $n$. In the general
case these parameters depend in a complicated way on the NP and matter
properties. For uncharged, unpolarized and isotropic matter, parameters
$\varepsilon$ are connected with $c$ parameters in a simple way
(see Eq.(\ref{E219}) below). For example, in the frame of the model
which we consider (Eqs.(\ref{E23}),(\ref{E24}), see also \cite{heavy}),
the effective NP Hamiltonian given by Eq.(\ref{E217}) can be determined
and, neglecting second order terms in $\delta\Lambda$, is equal to:
\begin{eqnarray}
\mathbf{H}^{NP} & = & \frac{1}{2E_{\nu}}(-A_{e}[E(1)\delta\Lambda+\delta\Lambda^{\dagger}E(1)]\label{E218}\\
 &  & +A_{n}[\delta\Lambda+\delta\Lambda^{\dagger}])\,,\nonumber \end{eqnarray}
 where $A_{n}=\sqrt{2}G_{F}N_{n}E_{\nu}$ depends on the neutron matter
density $N_{n}$ (in the Earth matter $A_{n}/A_{e}\approx1/2$) and
$E(1)_{\alpha\beta}=\delta_{\alpha e}\delta_{\beta e}$. Now, by comparing
equations Eq.(\ref{E217}) and Eq.(\ref{E218}), we can find the following
connection between their parameters ($\beta\geq\alpha$, $\chi_{\alpha\alpha}\equiv0$,
it may often be the case that $\delta\lambda_{e\mu}=\delta\lambda_{e\tau}=0$,
and then $c_{\alpha\beta}$ entirely describe the NP effects for modified
matter oscillations, compare Eq.(\ref{E3022}) below): \begin{equation}
\varepsilon_{\alpha\beta}e^{i\chi_{\alpha\beta}}=(\frac{A_{n}}{A_{e}}-\delta_{\alpha e}\delta_{\beta e})\, c_{\alpha\beta}-\delta_{\alpha e}(1-\delta_{\beta e})\,\delta\lambda_{e\beta}\,.\label{E219}\end{equation}
 For high energy neutrino beams with $E_{\nu}\approx O(GeV)$, the
following two small parameters, which describe $\nu SM$ neutrino
oscillations, are important (for lower energies, $E_{\nu}\approx O(MeV)$,
a third small factor, $A_{e}/\delta m_{31}^{2}$, would enter into
the game, too): \begin{equation}
\alpha=\frac{\delta m_{21}^{2}}{\delta m_{31}^{2}}\approx\pm0.03\,\,\,\,\text{and}\,\,\sin^{2}(2\theta_{13})\leq0.05\,.\label{E220}\end{equation}
 Also all $c_{\alpha\beta}$ and $\varepsilon_{\alpha\beta}$ parameters,
which describe NP, are small. Therefore, we can expand neutrino oscillation
probabilities in these small quantities, keeping only the leading
first order terms. In this approximation, the full transition probability,
for any flavor and the baseline $L$, can be decomposed into two terms,
the $\nu SM$ probability, and the correction to it given by NP: \begin{equation}
P_{P(\alpha)\rightarrow D(\beta)}(L)\,\,=\,\, P_{\alpha\rightarrow\beta}^{SM}(L)\,\,+\,\,\delta P_{\alpha\rightarrow\beta}^{NP}(L)\,.\label{E221}\end{equation}
 The NP correction probability we decompose again into two terms,
the $c$ term which is responsible for the initial and final neutrino
state modifications (see Appendix~\ref{app:approx}) and the $\varepsilon$
term which takes into account the NP influence on the coherent neutrino
scattering in matter: \begin{equation}
\delta P_{\alpha\rightarrow\beta}^{NP}(L)\,\,=\,\,\delta P_{\alpha\rightarrow\beta}^{c}(L)\,\,+\,\,\delta P_{\alpha\rightarrow\beta}^{\varepsilon}(L)\,.\label{E222}\end{equation}

The $\varepsilon$ term consists of two terms also. The first one,
which is responsible for an effective neutrino mass change, and the
second one, which describes the additional NP impact on coherent neutrino
scattering with matter particles: \begin{equation}
\delta P_{\alpha\rightarrow\beta}^{\varepsilon}(L)\,\,=\,\,\delta P_{\alpha\rightarrow\beta}^{mass}(L)\,\,+\,\,\delta P_{\alpha\rightarrow\beta}^{int}(L)\,.\label{E223}\end{equation}
 Generally, the production and detection states are not orthogonal:
\begin{equation}
\langle\nu_{P(\alpha)}|\nu_{D(\beta)}\rangle\neq\delta_{\alpha\beta}\,,\label{E224}\end{equation}
 and as a consequence the probability of neutrino oscillation is not
conserved: \begin{equation}
\sum_{all\,\beta}P_{P(\alpha)\rightarrow P(\beta)}\neq1\,.\label{E225}\end{equation}
 However, the neutrino oscillation probability of the $\nu SM$ is
normalized to one: \begin{eqnarray}
\sum_{all\,\beta}P_{\alpha\rightarrow\beta}^{SM}=1\,,\label{226a}\end{eqnarray}
 whereas the other terms satisfy: \begin{eqnarray}
\sum_{all\beta}\delta P_{\alpha\rightarrow\beta}^{c} & \neq & 0\,,\nonumber \\
\label{E226b}\\\sum_{all\beta}\delta P_{\alpha\rightarrow\beta}^{mass} & = & 0\,,\,\,\,\,\,\,\sum_{all\,\beta}\delta P_{\alpha\rightarrow\beta}^{int}=0\,.\nonumber \end{eqnarray}

In the next Chapter neutrino propagation in the Earth matter will
be discussed. In our numerical calculations of neutrino flavor transitions
we use the realistic PREM~I \cite{PREM-I} Earth density profile
model. However, explicit analytical formulas for the flavor transition
probabilities can only be given for the case of constant density matter.
Both NP corrections, which are important for neutrino transitions
in matter, the $\delta P_{\beta\rightarrow\gamma}^{int}(L)$ and the
$\delta P_{\beta\rightarrow\gamma}^{mass}(L)$, are small, therefore
their linear decomposition in terms of $\alpha$ and $sin(2\theta_{13})$
is a very good approximation (in all formulas, we assume $\delta m_{21}^{2}=\delta m_{sol}^{2}$,
and $\delta m_{31}^{2}=\pm\delta m_{atm}^{2}+\delta m_{sol}^{2}/2$,
where the upper/lower sign refers to the normal/inverted mass hierarchy
\cite{LMA}): \begin{equation}
\delta P_{\beta\rightarrow\gamma}^{int}\,=\, B_{\beta\gamma}^{0}\,+\,\,\alpha\, B_{\beta\gamma}^{\alpha}\,+\,\sin(2\theta_{13})\, B_{\beta\gamma}^{s}\,,\label{E227}\end{equation}
 and similarly \begin{equation}
\delta P_{\beta\rightarrow\gamma}^{mass}\,=\,\, C_{\beta\gamma}^{0}\,+\,\alpha\, C_{\beta\gamma}^{\alpha}\,+\,\,\sin(2\theta_{13})\, C_{\beta\gamma}^{s}\,.\label{E228}\end{equation}

The largest terms which are not suppressed neither by $\alpha$ nor
by $\sin(2\theta_{13})$, namely the terms $B_{\beta\gamma}^{0}$
and $C_{\beta\gamma}^{0}$, do not appear in any $\nu_{e}$ nor $\nu_{\bar{e}}$
related channels. Such terms are only present in $\nu_{\mu}\rightarrow\nu_{\tau}$,
$\nu_{\mu}\rightarrow\nu_{\mu}$, and $\nu_{\tau}\rightarrow\nu_{\tau}$
oscillation channels (and in the corresponding antineutrino oscillation
channels). For all such channels they are, up to the sign, the same
and have the following form: \begin{eqnarray}
B^{0} & = & \widehat{A}_{e}\,\sin(4\theta_{23})\{\sin(2\theta_{23})\,(\varepsilon_{\mu\mu}-\varepsilon_{\tau\tau})\label{E229}\\
 &  & +\,2\cos(2\theta_{23})\,\cos(\chi_{\mu\tau})\,\varepsilon_{\mu\tau}\}\sin^{2}(\Delta)\,,\nonumber \end{eqnarray}
 and \begin{eqnarray}
C^{0} & = & \widehat{A}_{e}\,\Delta\,\sin^{2}(2\theta_{23})\{-\cos(2\theta_{23})\,(\varepsilon_{\mu\mu}-\varepsilon_{\tau\tau})\label{E230}\\
 &  & +\,2\,\sin(2\theta_{23})\,\cos(\chi_{\mu\tau})\,\varepsilon_{\mu\tau}\}\sin(2\Delta)\,,\nonumber \end{eqnarray}
 where \begin{equation}
\widehat{A}_{e}\,\equiv\,\frac{A_{e}}{\delta m_{31}^{2}}\,\,\text{and}\,\,\Delta=\frac{\delta m_{31}^{2}L}{2E_{\nu}}\,.\label{E231}\end{equation}
 For the $\nu_{\mu}\rightarrow\nu_{\tau}$ and $\nu_{\mu}\rightarrow\nu_{\mu}$
transitions, we obtain: \begin{equation}
B_{\mu\tau}^{0}=-B_{\mu\mu}^{0}=B^{0}\,\,\text{and}\,\, C_{\mu\tau}^{0}=-C_{\mu\mu}^{0}=C^{0}\,.\label{E232}\end{equation}
 Unfortunately, these channels, wherein we can expect the largest
NP effects, are not easily achieved experimentally. To see the NP
effects in channels where they are suppressed, in the next Chapter,
we discuss one of the easiest experimental channels, the $\nu_{\mu}\rightarrow\nu_{e}$
one. All non-leading terms in Eq.(\ref{E227}) and Eq.(\ref{E228}),
together with all terms for antineutrino and time reversal channels,
are given in the Appendix~\ref{app:approx}.

In the model which we discuss, the $c$ parameters are constrained
from the existing experimental data (see \cite{heavy},\cite{parameters}):
\begin{eqnarray}
 & c_{ee}\leq0.0054\,,\,\,\, c_{\mu\mu}\leq0.0096\,,\,\,\, c_{\tau\tau}\leq0.016\,,\nonumber \\
 & |c_{e\mu}|=|c_{\mu e}|\leq0.0001\,,\,\,\,|c_{e\tau}|=|c_{\tau e}|\leq0.009\,,\label{E233}\\
 & |c_{\mu\tau}|=|c_{\tau\mu}|\leq0.012\,.\nonumber \end{eqnarray}
 There are no constraints on the phases. The $\varepsilon$ parameters
for the neutrino propagation in matter (Eq.(\ref{E217})) are determined
from relations given by Eq.(\ref{E219}).

These are also the confinements which we use in the next Chapter,
where we present some results of our numerical calculations of the
differences of the transition probabilities between CP-conjugate channels:\begin{equation}
\Delta P_{\alpha\rightarrow\beta}^{CP}(L)\,\,=\,\, P_{\alpha\rightarrow\beta}(L)\,\,-\,\, P_{\overline{\alpha}\rightarrow\overline{\beta}}(L)\,.\label{E234}\end{equation}

\section{New Physics in Future Neutrino Oscillation Experiments}

\label{sec:sims}In order to check the effects of the NP described
above, the probability differences $\Delta P_{\alpha\rightarrow\beta}^{CP}$
(Eq.(\ref{E234})) for two energy ranges and several baselines, have
been calculated. Both, the energy ranges and the baselines have been
chosen in prospect of the existing, planned, and feasible experiments.
In view of $Beta\, Beam$ and $Super\, Beam$ experiments, the first
energy range is $E_{\nu}=0.1\div5\, GeV$ and the considered baselines
are $L=130\,,\,295\,,\,810\, km$ (\cite{SBBeam}). In view of $Neutrino\, Factory$
experiments, the second energy range is $E_{\nu}=1\div50\, GeV$ and
the considered baselines are $L=732\,,\,3000\,,\,7500\, km$ (\cite{NuFact}).
In our numerical calculations of neutrino flavor transitions, we use
the realistic PREM~I \cite{PREM-I} Earth density profile model,
which assesses the actual matter density $\rho$ and the actual electron
fraction $Y_{e}$, along the neutrino flight path in the Earth's interior.
Then: \begin{eqnarray}
A_{e}\,[eV^{2}] & = & 7.63\times10^{-5}\,[\frac{\rho}{g/cm^{3}}]\,[\frac{Y_{e}}{0.5}]\,[\frac{E_{\nu}}{GeV}]\,,\nonumber \\
\label{E301}\\A_{n}\,[eV^{2}] & = & 7.63\times10^{-5}\,[\frac{\rho}{g/cm^{3}}]\,[1-Y_{e}]\,[\frac{E_{\nu}}{GeV}]\,.\nonumber \end{eqnarray}

Note here that, for $L\lesssim874\, km$ neutrinos pass only the first
shell of the Earth's crust, with a constant density $\rho=2.6\, g/cm^{3}$
and $Y_{e}=0.494$, thus $A_{e}[eV^{2}]=1.96\times10^{-4}[E_{\nu}/GeV]$
and $A_{n}[eV^{2}]=1.0\times10^{-4}[E_{\nu}/GeV]$.

The $\nu SM$ oscillation parameters, together with their $\pm2\sigma$
errors ($95\%$ C.L., correlations among parameters are currently
considered small), are taken from the current global best fit values
\cite{LMA}: \begin{eqnarray}
\sin^{2}(\theta_{13}) & = & 0.9_{-0.9}^{+2.3}\times10^{-2}\,,\nonumber \\
\delta m_{sol}^{2} & = & 7.92\,(1\pm0.09)\times10^{-5}\,[eV^{2}]\,,\nonumber \\
\sin^{2}(\theta_{12}) & = & 0.314\,(1_{-0.15}^{+0.18})\,,\label{E302}\\
\delta m_{atm}^{2} & = & 2.4\,(1_{-0.26}^{+0.21})\times10^{-3}\,[eV^{2}]\,,\nonumber \\
\sin^{2}(\theta_{23}) & = & 0.44\,(1_{-0.22}^{+0.41})\,.\nonumber \end{eqnarray}

In order to implement the effects of heavy neutrinos, as discussed
in the previous Chapter, the matrix $V$ must be introduced. As the
number of heavy non-decoupling neutrinos is unknown, we parametrize
it in a simplified way, using a single {}``effective'' heavy neutrino
state. In this way, the number of independent quantities parameterizing
the matrix $\mathcal{U}_{\ell\, i}$ in Eq.(\ref{E27}) are $6$ moduli
and $3$ CP phases: $3$ standard mixing angles ($\theta_{12},\theta_{13},\theta_{23}$)
+ $1$ standard Dirac phase ($\delta_{13}$) + $3$ new small NP mixing
angles ($\theta_{14},\theta_{24},\theta_{34}$) + $2$ new NP Dirac
phases ($\delta_{24},\delta_{34}$). Then \begin{equation}
V=\left(\begin{array}{c}
\sin(\theta_{14})\\
\cos(\theta_{14})\,\sin(\theta_{24})\, e^{-i\delta_{24}}\\
\cos(\theta_{14})\,\cos(\theta_{24})\,\sin(\theta_{34})\, e^{-i\delta_{34}}\end{array}\right)\,,\label{E3021}\end{equation}
 and \begin{widetext} \begin{equation}
\delta\Lambda=\left(\begin{array}{ccc}
1-\cos(\theta_{14}) & 0 & 0\\
\sin(\theta_{14})\,\sin(\theta_{24})\, e^{-i\delta_{24}} & 1-\cos(\theta_{24}) & 0\\
\sin(\theta_{14})\,\cos(\theta_{24})\,\sin(\theta_{34})\, e^{-i\delta_{34}} & \sin(\theta_{24})\,\sin(\theta_{34})\, e^{-i(\delta_{34}-\delta_{24})} & 1-\cos(\theta_{34})\end{array}\right)\,.\label{E3022}\end{equation}
 \end{widetext}

Two sets of $V$ parameters, both satisfying present experimental
constraints given by Eq.(\ref{E233}) are discussed below: \begin{equation}
\mathrm{(A)\,:}\,\,\,\left(\begin{array}{c}
0.001\\
0.1\, e^{-i\delta_{24}}\\
0.1\, e^{-i\delta_{34}}\end{array}\right)\,,\,\,\,\,\,\mathrm{(B)\,:}\,\,\,\left(\begin{array}{c}
0.01\\
0.01\, e^{-i\delta_{24}}\\
0.1\, e^{-i\delta_{34}}\end{array}\right)\,.\label{E303}\end{equation}

All calculations have been performed assuming the direct mass scheme
only. As nothing is known about values of CP phases, we allow them
to vary freely.

For both $V$ sets, we notice that, the biggest potential of the discovery
of the possible presence of any NP is pronounced in oscillation channels
in which $\nu_{e}$, $\nu_{\bar{e}}$ are not involved at all, that
is in $\nu_{\mu}\rightarrow\nu_{\tau}$ and $\nu_{\mu}\rightarrow\nu_{\mu}$
(including the corresponding antineutrino channels). The effects are
especially visible for two baselines, $L=3000\, km$ and $L=7500\, km$,
which, for other reasons, are also considered {}``magic'' for future
$Neutrino\, Factory$ experiments. Moreover, comparing numerical results
for these two sets of $V$ parameters, we can clearly see that, as
in Eq.(\ref{E303}.B) the magnitude of the middle row is $10$ times
smaller than the corresponding magnitude in Eq.(\ref{E303}.A), the
NP effects for the second $V$ set Eq.(\ref{E303}.B) are smaller
by a similar factor (compare Eq.(\ref{E304}) below), too.

In order to find how the uncertainty of the estimation of the $\nu SM$
oscillation parameters can mimic any possible NP effects, we have
also performed calculations allowing all $\nu SM$ oscillation parameters
to vary by $\pm3\%$ and $\pm2\sigma$ (note here that, the $\pm2\sigma$
test is also useful in qualitative estimation of the effect of the
uncertainties in the Earth density profile on the $\nu SM$ results).
We have found that in general, in order to give a chance for a discovery
of NP effects, it is required that $\nu SM$ neutrino oscillation
parameter errors should be diminished to the values expected in the
future (of about $\pm3\%$). However, using neutrinos with energies
$E_{\nu}\gtrsim15\, GeV$, together with the {}``magic'' baselines
mentioned above, it should even be possible to see NP effects with
today's $\nu SM$ neutrino oscillation parameter uncertainties (of
about $\pm2\sigma$). But, this chance depends on the actual magnitudes
of the NP parameters and the actual precision of the experiment (which
may, of course, be required to be much better than the allowed $\nu SM$
neutrino oscillation parameter errors).

\begin{figure*}
\includegraphics[width=17.5cm,height=19.5cm]{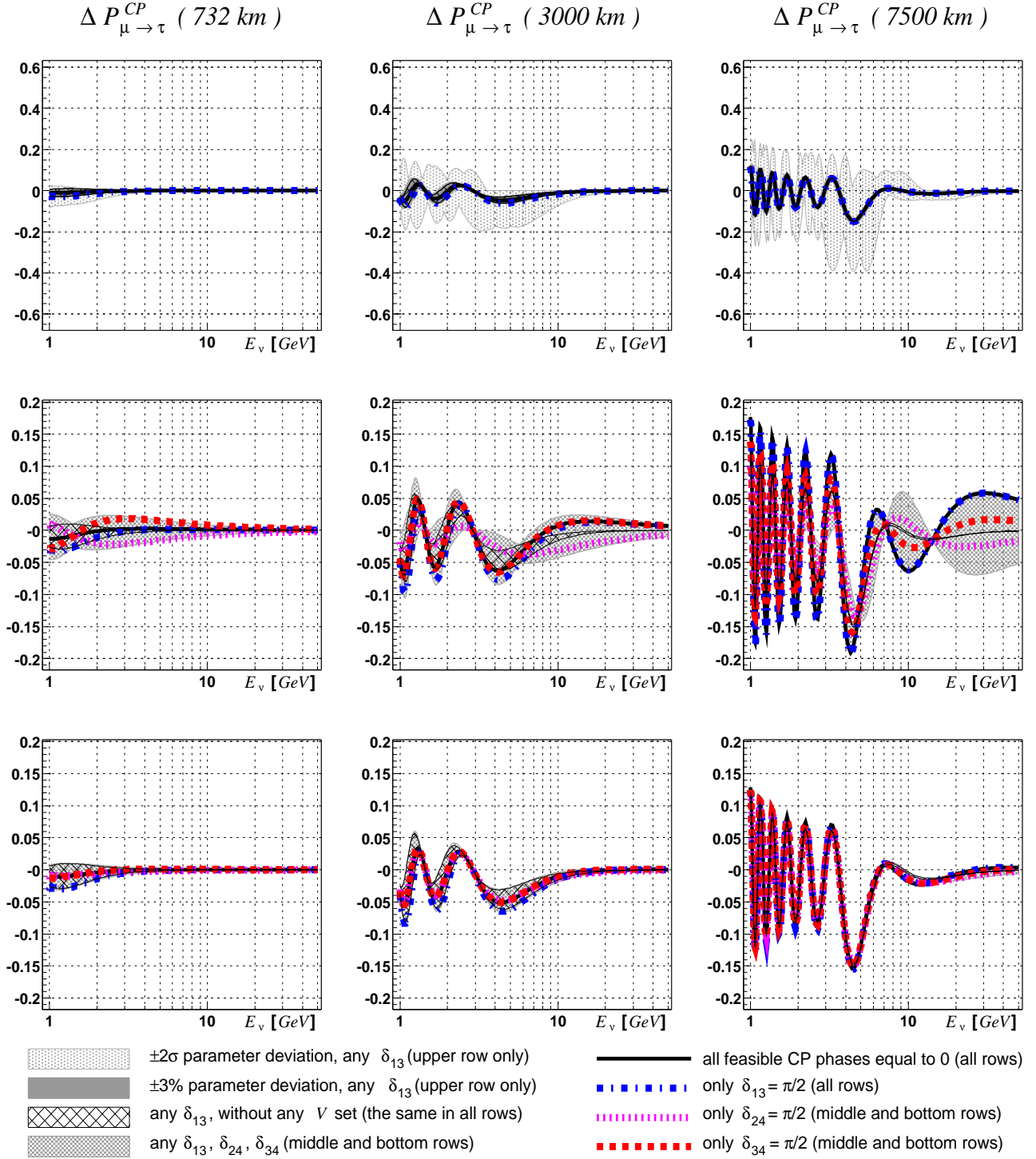}

\caption{\label{F1}(Color online) The probability differences $\Delta P_{\mu\rightarrow\tau}^{CP}$
for three baselines $L=732\,,\,3000\,,\,7500\, km$ (each column of
graphs corresponds to a single $L$), for the energy range $E_{\nu}=1\div50\, GeV$.
Calculations assuming the $\nu SM$ only are presented in the upper
row of graphs, whereas results with NP are shown in the middle and
bottom rows, except for the hashed band, identical in all rows, which
corresponds to the current global best fit parameters with any feasible
$\delta_{13}$ value and $\nu SM$ only (no NP). The dark (light)
gray band in the upper row corresponds to $\pm3\%$ ($\pm2\sigma$)
deviations of the $\nu SM$ neutrino oscillation parameters with any
feasible $\delta_{13}$ value. The light gray band in the middle (bottom)
row corresponds to the current global best fit parameters with NP
$V$ parameters set Eq.(\ref{E303}.A) (Eq.(\ref{E303}.B)), with
any feasible values of $\delta_{13},\delta_{24},\delta_{34}$. Curves
present in all graphs correspond to the current global best fit parameters
with all feasible CP phases equal to $0$ (solid curves) and with
exactly one of them equal to $\pi/2$ (dotted and dashed curves).}
\end{figure*}

\begin{figure*}
\includegraphics[width=17.5cm,height=19.5cm]{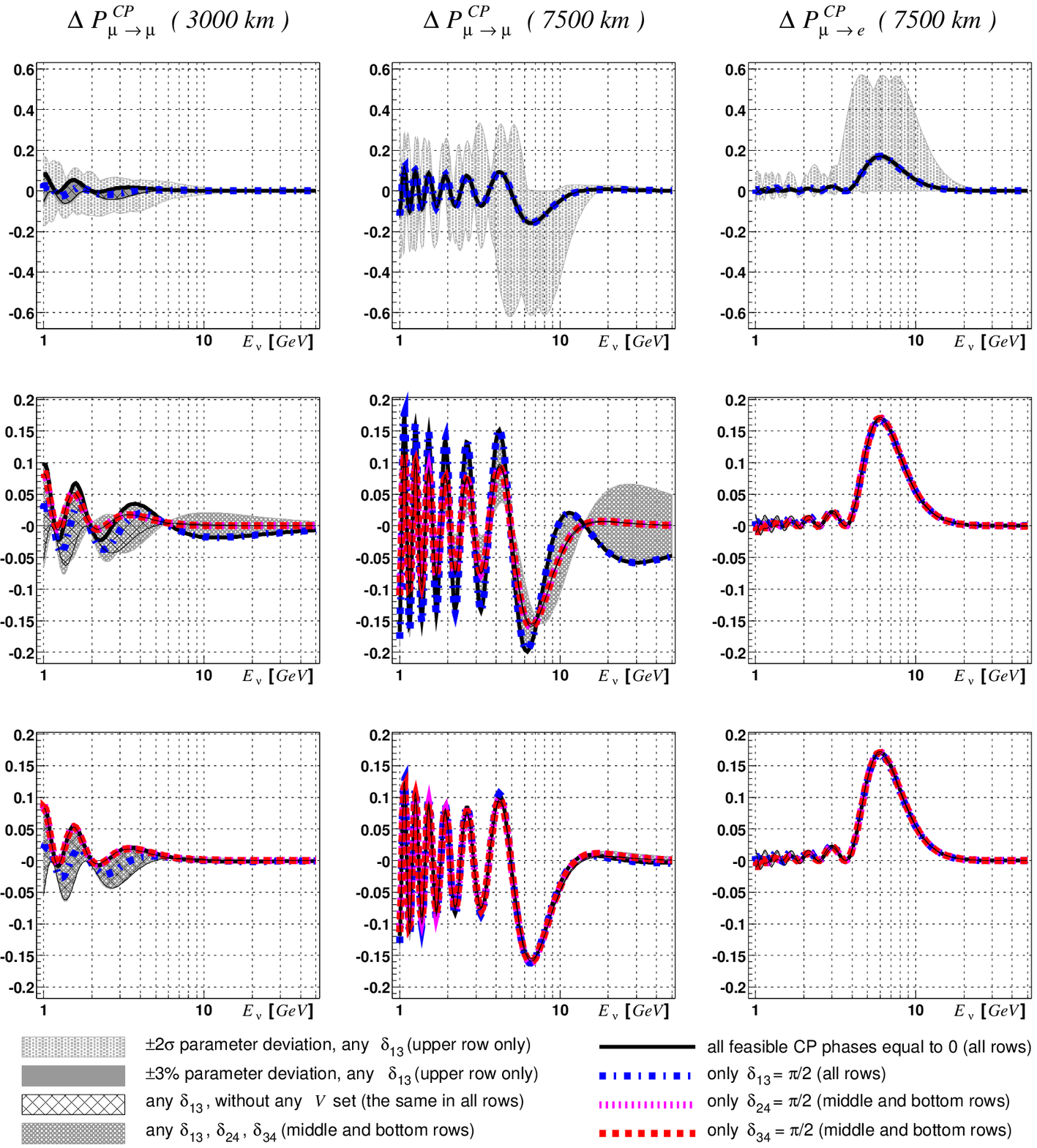}

\caption{\label{F2}(Color online) The probability differences $\Delta P_{\mu\rightarrow\mu}^{CP}$
for two baselines $L=3000\,,\,7500\, km$ (the left and middle columns
of graphs), and $\Delta P_{\mu\rightarrow e}^{CP}$ for $L=7500\, km$
(the right column of graphs) for the energy range $E_{\nu}=1\div50\, GeV$.
Calculations assuming the $\nu SM$ only are presented in the upper
row of graphs, whereas results with NP are shown in the middle and
bottom rows, except for the hashed band, identical in all rows, which
corresponds to the current global best fit parameters with any feasible
$\delta_{13}$ value and $\nu SM$ only (no NP). The dark (light)
gray band in the upper row corresponds to $\pm3\%$ ($\pm2\sigma$)
deviations of the $\nu SM$ neutrino oscillation parameters with any
feasible $\delta_{13}$ value. The light gray band in the middle (bottom)
row corresponds to the current global best fit parameters with NP
$V$ parameters set Eq.(\ref{E303}.A) (Eq.(\ref{E303}.B)), with
any feasible values of $\delta_{13},\delta_{24},\delta_{34}$. Curves
present in all graphs correspond to the current global best fit parameters
with all feasible CP phases equal to $0$ (solid curves) and with
exactly one of them equal to $\pi/2$ (dotted and dashed curves).}
\end{figure*}

In Fig.\ref{F1}, the probability differences $\Delta P_{\mu\rightarrow\tau}^{CP}$,
for the second energy range, and the corresponding baselines set,
are shown. The NP effects at $L=732\, km$, when scaled by a factor
$L/E_{\nu}$, can be used as an approximate estimation of the expected
results for the first energy range with its considered baselines (for
which we do not show pictures in this paper).

In the left and middle columns of Fig.\ref{F2}, the probability differences
$\Delta P_{\mu\rightarrow\mu}^{CP}$ for the second energy range and
the two {}``magic'' $Neutrino\, Factory$ baselines, are shown.
For this oscillation channel, the NP effect at $L=732\, km$ is significantly
smaller than in the $\nu_{\mu}\rightarrow\nu_{\tau}$ channel at the
same distance. This is also the case for this oscillation channel
in the first energy range with its considered baselines. Hence for
this reason, we do not show the corresponding pictures in this paper
either.

Note that, in general, the transition probabilities of both, neutrino
and antineutrino, oscillations depend on matter properties in a different
way. However, at higher energies the $\nu SM$ dependence is very
much similar, thus any possible uncertainties in the Earth density
profile cancel in the $\nu SM$ probability differences (see the upper
row of graphs, in both figures, for $E_{\nu}\gtrsim15\, GeV$). In
this way, any signal that the probability differences in this energy
range are distinctly different from zero will indicate that some NP
exists. The statistical significance of such signal depends on its
actual magnitude. For example, for the first $V$ set Eq.(\ref{E303}.A),
and $L=7500\, km$, there exists a maximum at $E_{\nu}\approx30\, GeV$,
in which $\Delta P_{\mu\rightarrow\tau}^{CP}\approx-\Delta P_{\mu\rightarrow\mu}^{CP}\approx\pm0.06$,
while $P_{\mu\rightarrow\tau}\approx P_{\mu\rightarrow\mu}\approx0.5$,
and thus $\Delta P^{CP}/P\approx\pm0.12$ (see the middle row of graphs
in both figures). The neutrino energy dependent magnitude of this
effect can well be reproduced just by the largest term in the $\delta P_{\beta\rightarrow\gamma}^{mass}$
NP correction (responsible for the effective neutrino mass change,
see Eq.(\ref{E228})), which is not suppressed neither by $\alpha$
nor by $\sin(2\theta_{13})$. Taking into account the fact that $P_{\overline{\alpha}\rightarrow\overline{\beta}}(\delta_{ij},\,\widehat{A}_{e})=P_{\alpha\rightarrow\beta}(-\delta_{ij},\,-\widehat{A}_{e})$,
one can write (see Eq.(\ref{E230})): \begin{eqnarray}
\Delta P_{\mu\rightarrow\tau}^{CP} & \approx & -\Delta P_{\mu\rightarrow\mu}^{CP}\,\approx\,2\, C^{0}\label{E304}\\
 & \approx & 4\,\widehat{A}_{e}\,\sin^{3}(2\theta_{23})\,\Delta\,\sin(2\Delta)\,\cos(\chi_{\mu\tau})\,\varepsilon_{\mu\tau}\,.\nonumber \end{eqnarray}
 It should be noted, however, that the above formula does not reproduce
the $\Delta P_{\mu\rightarrow\tau}^{CP}$ well, in case one of the
NP feasible CP phases ($\chi_{\mu\tau}=\delta_{34}-\delta_{24}$)
is equal to $\pi/2$ (see the dotted and dashed curves in Fig.\ref{F1}).
The reason is that in the above estimation we completely neglect several
explicit CP asymmetry bracing terms, for example proportional to $\sin(\chi_{\alpha\beta})$,
which are relevant in the $\Delta P_{\mu\rightarrow\tau}^{CP}$ case,
but which never appear in the case of $\Delta P_{\mu\rightarrow\mu}^{CP}$
(see the Appendix~\ref{app:approx}, note also that, these asymmetry
terms become dominant at short baselines, and that is why the NP effect
at $L=732\, km$ is significantly smaller in the $\nu_{\mu}\rightarrow\nu_{\mu}$
channel than in the $\nu_{\mu}\rightarrow\nu_{\tau}$ channel at the
same distance). From the above formula we can easily learn that, the
magnitude of this NP effect is linearly proportional to both, the
actual matter density through the term $\widehat{A}_{e}$, and to
the $\varepsilon_{\mu\tau}$ NP parameter (thus, as the value of the
$\varepsilon_{\mu\tau}$ parameter that results from the Eq.(\ref{E303}.B)
is $10$ times smaller than the one from the Eq.(\ref{E303}.A), the
NP effect for the second $V$ set Eq.(\ref{E303}.B) is smaller by
a similar factor, too).

As already mentioned, one of the easiest channels, from the experimental
point of view, is the $\nu_{\mu}\rightarrow\nu_{e}$ one, but it will
be difficult to observe any NP there (as will be in all channels in
which $\nu_{e}$ or $\nu_{\bar{e}}$ are involved). In the right column
of Fig.\ref{F2}, the probability differences $\Delta P_{\mu\rightarrow e}^{CP}$
for the second energy range and the longest considered baseline, are
shown (where the biggest effects are expected). It can be seen that
the NP effect is rather miserable in this oscillation channel, regardless
of the $V$ set given by Eq.(\ref{E303}). This conclusion holds also
for all another similar oscillation channels and both energy ranges
with the corresponding baselines. Moreover, these oscillation channels
are sensitive to the (not so very well known) value of the $\sin^{2}(\theta_{13})$.
In the other channels, without $\nu_{e}$ and $\nu_{\bar{e}}$, the
dependence on the $\sin^{2}(\theta_{13})$ is small, giving a better
chance to see the NP effects.

\section{Conclusions}

\label{sec:fruits} In the paper, by introducing a mixing of the low
mass {}``active'' neutrinos with heavy ones, we have investigated
a possible New Physics (NP) scenario which is already present at the
$TeV$ scale, that means, at energies close to our present-day experimental
facilities. In the presented model (as also in any model with lepton
flavor violation), the effective mixing matrix is non-unitary, resulting
in non-orthogonal neutrino production and detection states. This leads
to the modification of the neutrino oscillations in vacuum. Additionally,
non-standard neutrino interactions with matter particles influence
oscillation effects also. First order approximation formulas for the
flavor transition probabilities, in constant density matter, for all
experimentally available channels, have been given. The possibilities
of experimental verification of such model predictions have been discussed
in prospect of the existing, planned, and feasible $Beta\, Beam$,
$Super\, Beam$, and $Neutrino\, Factory$ experiments. Numerical
calculations of flavor transition probabilities for two sets (satisfying
present experimental constraints) of NP parameters which describe
a single {}``effective'' heavy neutrino state have been performed.
They took into account two energy ranges and several baselines, assuming
both the current ($\pm2\sigma$) and the expected in future ($\pm3\%$)
errors of today's $\nu SM$ neutrino oscillation parameters, keeping
unchanged their present central values. The realistic PREM~I Earth
density profile model has been applied. One of the easiest channels,
from the experimental point of view, is the $\nu_{\mu}\rightarrow\nu_{e}$
one, but it will be difficult to observe any NP there (as it will
be also in all channels in which $\nu_{e}$ or $\nu_{\bar{e}}$ are
involved). It appears that the biggest potential of the discovery
of any possible presence of the NP is pronounced in oscillation channels
in which $\nu_{e}$, $\nu_{\bar{e}}$ are not involved at all, that
is in $\nu_{\mu}\rightarrow\nu_{\tau}$ and $\nu_{\mu}\rightarrow\nu_{\mu}$
(and in the corresponding antineutrino channels). The effects are
especially visible for the two so called {}``magic'' $Neutrino\, Factory$
baselines, $L=3000\, km$ and $L=7500\, km$. We have also found that
in general, in order to give a chance for a discovery of the NP effects,
it is required that $\nu SM$ neutrino oscillation parameter errors
should be diminished to the values expected in the future (of about
$\pm3\%$). However, using neutrinos with energies $E_{\nu}\gtrsim15\, GeV$
together with the {}``magic'' baselines mentioned above, it should
even be possible with today's $\nu SM$ neutrino oscillation parameter
uncertainties (of about $\pm2\sigma$). But, this chance depends on
the actual magnitudes of the NP parameters and the actual precision
of the experiment (which may, of course, be required to be much better
than the allowed $\nu SM$ neutrino oscillation parameter errors).

Finally, it should be stressed that the full quantitative treatment
of the NP effects in future facilities should be based on realistic
observables related, for example, to the expected numbers of events
at the corresponding facilities. However, as stated in the beginning
of the Chapter~\ref{sec:theory}, this would require not only the
knowledge of the NP generated transition probability modifications,
but also the NP generated modifications of the production and detection
neutrino cross sections need to be known. Some preliminary studies
(see \cite{heavy}) suggest that the expected effects on the last
two terms can be of the same order as these on the first term (shown
in this paper).

\begin{acknowledgments}
This work has been supported by the Polish Ministry of Science under
Grant 1P03B04926. 
\end{acknowledgments}
\appendix
\begin{widetext}

\section{First order approximations for flavor transition probabilities}

\label{app:approx}Herein, we collect all formulas for the flavor
transition probabilities, in constant density matter, for all experimentally
available neutrino and antineutrino channels. As already mentioned,
in all formulas, we assume $\delta m_{21}^{2}=\delta m_{sol}^{2}$,
and $\delta m_{31}^{2}=\pm\delta m_{atm}^{2}+\delta m_{sol}^{2}/2$,
where the upper (lower) sign refers to the normal (inverted) mass
hierarchy \cite{LMA}.

Firstly, we show formulas required in order to calculate the $c$
term in Eq.(\ref{E222}), which is responsible for the initial and
final neutrino state modifications (see also \cite{Pramana}): \begin{eqnarray}
\delta P_{\alpha\rightarrow\beta}^{c}(L) & = & (c_{\alpha\alpha}+c_{\beta\beta})P_{\alpha\rightarrow\beta}^{SM}(L)-2\, Re\,(c_{\alpha\beta})\label{A00}\\
 &  & -\,4\,\sum_{i>k}Re\,[(\delta\widetilde{T}^{c})_{\alpha\beta}^{ik}]\,\sin^{2}(\frac{\widetilde{\Delta}_{ik}}{2})-2\,\sum_{i>k}Im\,[(\delta\widetilde{T}^{c})_{\alpha\beta}^{ik}]\,\sin(\widetilde{\Delta}_{ik})\,,\nonumber \end{eqnarray}
 where \begin{eqnarray}
(\delta\widetilde{T}^{c})_{\alpha\beta}^{ik} & = & -\sum_{\gamma}\,(\delta\lambda_{\alpha\gamma}\,\widetilde{U}_{\gamma i}\,\widetilde{U}_{\beta k}+\delta\lambda_{\beta\gamma}\,\widetilde{U}_{\alpha i}\,\widetilde{U}_{\gamma k})\,\widetilde{U}_{\alpha k}^{*}\,\widetilde{U}_{\beta i}^{*}\label{A01}\\
 &  & -\sum_{\gamma}\,(\delta\lambda_{\alpha\gamma}^{*}\,\widetilde{U}_{\gamma k}^{*}\,\widetilde{U}_{\beta i}^{*}+\delta\lambda_{\beta\gamma}^{*}\,\widetilde{U}_{\alpha k}^{*}\,\widetilde{U}_{\gamma i}^{*})\,\widetilde{U}_{\alpha i}\,\widetilde{U}_{\beta k}\,,\nonumber \end{eqnarray}
 and \begin{eqnarray}
\widetilde{\Delta}_{ik} & = & \frac{\delta\widetilde{m}_{ij}^{2}\, L}{2\, E_{\nu}}\,.\label{A02}\end{eqnarray}

Here, $\widetilde{U}_{\alpha i}$ and $\delta\widetilde{m}_{ij}^{2}$
are, respectively, the effective mixing matrix elements and the effective
mass square differences, coming from the diagonalization of the $\nu SM$
Hamiltonian in matter (without any NP, see, for example, \cite{EffMattParams}),
and $\delta\lambda_{\alpha\beta}$ can be calculated by putting $\delta\Lambda=\mathbf{1}-\mathcal{U}\, U^{\dagger}$
(compare Eq.(\ref{E29})).

Secondly, as described in Chapter~\ref{sec:theory}, not all available
channels were discussed in this paper. We were only interested in
$\nu_{\mu}\rightarrow\nu_{e}$, $\nu_{\mu}\rightarrow\nu_{\tau}$
and $\nu_{\mu}\rightarrow\nu_{\mu}$ channels, including the corresponding
antineutrino transitions. In this Appendix, however, we collect all
required equations.

In any case, the total transition probability is decomposed into several
terms according to Eqs.(\ref{E221}),(\ref{E222}),(\ref{E223}),(\ref{E227}),(\ref{E228}).
In general, the mass term $\delta P_{\beta\rightarrow\gamma}^{mass}$
(Eqs.(\ref{E223}),(\ref{E228})) does not vanish only for channels
in which $\nu_{e}$, $\nu_{\bar{e}}$ do not enter. This term is also,
up to the sign, the same for all such channels. Let us put: \begin{equation}
C^{\alpha}=\Delta\,\sin(2\Delta)\,\sin(2\theta_{12})\,\sin^{2}(2\theta_{23})\,\left[\cos(\theta_{23})\,\cos(\chi_{e\mu})\,\varepsilon_{e\mu}-\cos(\chi_{e\tau})\,\sin(\theta_{23})\,\varepsilon_{e\tau}\right]\,,\label{A3}\end{equation}
 \begin{equation}
C^{s}=\frac{\widehat{A}_{e}}{1-\widehat{A}_{e}}\biggl\{\Delta\,\sin(2\Delta)\,\sin^{2}(2\theta_{23})\,\left[\cos(\delta_{13}+\chi_{e\mu})\,\sin(\theta_{23})\,\varepsilon_{e\mu}+\cos(\theta_{23})\,\cos(\delta_{13}+\chi_{e\tau})\,\varepsilon_{e\tau}\right]\biggr\}\,.\label{A4}\end{equation}
 For the $\nu_{\mu}\rightarrow\nu_{e}$ transition, we obtain: \begin{equation}
\delta P_{\mu\rightarrow e}^{mass}\,=0\,,\,\,\,\, B_{\mu e}^{0}=0\,,\label{A5}\end{equation}
 and two non-vanishing terms have the following form: \begin{eqnarray}
B_{\mu e}^{\alpha} & = & \frac{1}{(\widehat{A}_{e}-1)\widehat{A}_{e}^{2}}\cos(\theta_{23})\sin(2\theta_{12})\biggl\{\varepsilon_{e\tau}\cos(\theta_{23})\sin(\theta_{23})\biggl[\cos(\chi_{e\tau})[\widehat{A}_{e}\cos(2\Delta)\label{A6}\\
 &  & -\widehat{A}_{e}\cos(2(\widehat{A}_{e}-1)\Delta)-2(\widehat{A}_{e}-2)\sin^{2}(\widehat{A}_{e}\Delta)]+\sin(\chi_{e\tau})\widehat{A}_{e}[\sin(2\Delta)\nonumber \\
 &  & -\sin(2(1-\widehat{A}_{e})\Delta)-\sin(2\widehat{A}_{e}\Delta)]\biggr]\nonumber \\
 &  & +\varepsilon_{e\mu}\biggl[2\cos(\chi_{e\mu})\sin(\widehat{A}_{e}\Delta)[2(\widehat{A}_{e}-1)\cos^{2}(\theta_{23})\sin(\widehat{A}_{e}\Delta)+\widehat{A}_{e}(\sin((\widehat{A}_{e}-2)\Delta)\nonumber \\
 &  & +\sin(\widehat{A}_{e}\Delta))\sin^{2}(\theta_{23})]+\sin(\chi_{e\mu})\widehat{A}_{e}\sin^{2}(\theta_{23})[\sin(2\Delta)-\sin(2(1-\widehat{A}_{e})\Delta)\nonumber \\
 &  & -\sin(2\widehat{A}_{e}\Delta)]\biggr]\biggr\}\,,\nonumber \end{eqnarray}
 \begin{eqnarray}
B_{\mu e}^{s} & = & \frac{1}{(\widehat{A}_{e}-1)\widehat{A}_{e}^{2}}\sin(\theta_{23})\biggl\{\varepsilon_{e\tau}\sin(\theta_{23})\cos(\theta_{23})\biggl[\cos(\delta_{13}+\chi_{e\tau})[1+\widehat{A}_{e}\label{A7}\\
 &  & -(\widehat{A}_{e}-1)\cos(2\Delta)-\cos(2(\widehat{A}_{e}-1)\Delta)-(1-\widehat{A}_{e})\cos(2\widehat{A}_{e}\Delta)]\nonumber \\
 &  & +\sin(\delta_{13}+\chi_{e\tau})(\widehat{A}_{e}-1)[\sin(2\Delta)-\sin(2(1-\widehat{A}_{e})\Delta)-\sin(2\widehat{A}_{e}\Delta)]\biggr]\nonumber \\
 &  & +\varepsilon_{e\mu}\biggl[\cos(\delta_{13}+\chi_{e\mu})[(\widehat{A}_{e}-1)\cos^{2}(\theta_{23})(\cos(2\Delta)-\cos(2(\widehat{A}_{e}-1)\Delta)\nonumber \\
 &  & +2\sin^{2}(\widehat{A}_{e}\Delta))+4\widehat{A}_{e}\sin^{2}((\widehat{A}_{e}-1)\Delta)\sin^{2}(\theta_{23})]\nonumber \\
 &  & +\sin(\delta_{13}+\chi_{e\mu})(\widehat{A}_{e}-1)\cos^{2}(\theta_{23})[\sin(2(1-\widehat{A}_{e})\Delta)-\sin(2\Delta)+\sin(2\widehat{A}_{e}\Delta)]\biggr]\biggr\}\,.\nonumber \end{eqnarray}
 For the $\nu_{\mu}\rightarrow\nu_{\tau}$ transition, all terms are
different from zero. According to the previous discussion, the leading
term $B_{\mu\tau}^{0}=B^{0}$ (see Eq.(\ref{E232})) and the two non-leading
terms are given by: \begin{eqnarray}
B_{\mu\tau}^{\alpha} & = & \frac{\sin(2\,\theta_{23})}{2\,\left(-1+\widehat{A}_{e}\right)\,\widehat{A}_{e}}\,\biggl\{\varepsilon_{e\mu}\,\sin(2\,\theta_{12})\,\sin(\theta_{23})\,\biggl[\cos(\chi_{e\mu})\,\biggl[\left[\cos(2\,(-1+\widehat{A}_{e})\,\Delta)\right.\label{A8}\\
 &  & \left.-\cos(2\,\widehat{A}_{e}\,\Delta)\right]\left(1-\widehat{A}_{e}+\cos(2\,\theta_{23})\right)\,+2\,(2-\widehat{A}_{e}-2\,\widehat{A}_{e}^{2})\,\cos^{2}(\theta_{23})\,\sin^{2}(\Delta)\nonumber \\
 &  & +(4\,\widehat{A}_{e}^{2}-2\,\widehat{A}_{e})\,\sin^{2}(\theta_{23})\,\sin^{2}(\Delta)\biggr]-\widehat{A}_{e}\,\sin(\chi_{e\mu})\,\left[\sin(2\,\Delta)+\sin(2\,(1-\widehat{A}_{e})\,\Delta)\right.\nonumber \\
 &  & \left.+\sin(2\,\widehat{A}_{e}\,\Delta)\right]\biggr]\nonumber \\
 &  & +\varepsilon_{e\tau}\cos(\theta_{23})\sin(2\,\theta_{12})\,\biggl[\cos(\chi_{e\tau})\,\biggl[\left[\cos(2\,(-1+\widehat{A}_{e})\,\Delta)-\cos(2\,\widehat{A}_{e}\,\Delta)\right]\,\nonumber \\
 &  & \left(-1+\widehat{A}_{e}+\cos(2\,\theta_{23})\right)+2\,\widehat{A}_{e}\,\cos^{2}(\theta_{23})\,\sin^{2}(\Delta)-4\,\widehat{A}_{e}^{2}\,\cos(2\,\theta_{23})\,\sin^{2}(\Delta)\nonumber \\
 &  & +2\,(\widehat{A}_{e}-2)\,\sin^{2}(\Delta)\,\sin^{2}(\theta_{23})\biggr]+\widehat{A}_{e}\,\sin(\chi_{e\tau})\,\left[-\sin(2\,\Delta)+\sin(2\,(1-\widehat{A}_{e})\,\Delta)\right.\nonumber \\
 &  & \left.+\sin(2\,\widehat{A}_{e}\,\Delta)\right]\biggr]\,\nonumber \\
 &  & -4\,\left(-1+\widehat{A}_{e}\right)\,\widehat{A}_{e}^{2}\,\cos^{2}(\theta_{12})\,\cos(2\,\theta_{23})\,\left(2\,\Delta\,\cos(\Delta)-\sin(\Delta)\right)\,\sin(\Delta)\,\nonumber \\
 &  & \biggl[\sin(2\,\theta_{23})\,(\varepsilon_{\mu\mu}-\varepsilon_{\tau\tau})+2\,\cos(2\,\theta_{23})\,\cos(\chi_{\mu\tau})\,\varepsilon_{\mu\tau}\biggr]\biggr\}\,,\nonumber \end{eqnarray}
 \begin{eqnarray}
B_{\mu\tau}^{s} & = & \frac{\sin(2\,\theta_{23})}{2\,\left(-1+\widehat{A}_{e}\right)^{2}}\,\biggl\{\varepsilon_{e\mu}\,\cos(\theta_{23})\,\biggl[2\,\cos(\delta_{13}+\chi_{e\mu})\,\sin(\Delta)\,\label{A9}\\
 &  & \left[\left(-\widehat{A}_{e}-(1-4\,\widehat{A}_{e}+2\,\widehat{A}_{e}^{2})\,\cos(2\,\theta_{23})\right)\sin(\Delta)-\left(\widehat{A}_{e}-\cos(2\,\theta_{23})\right)\sin(\Delta-2\,\widehat{A}_{e}\,\Delta)\right]\nonumber \\
 &  & +\left(-1+\widehat{A}_{e}\right)\,\sin(\delta_{13}+\chi_{e\mu})\,\left[\sin(2\,\Delta)-\sin(2\,(1-\widehat{A}_{e})\,\Delta)-\sin(2\,\widehat{A}_{e}\,\Delta)\right]\biggr]\nonumber \\
 &  & +\varepsilon_{e\tau}\sin(\theta_{23})\,\biggl[2\,\cos(\delta_{13}+\chi_{e\tau})\sin(\Delta)\left[\left(-\widehat{A}_{e}+(1-4\,\widehat{A}_{e}+2\,\widehat{A}_{e}^{2})\,\cos(2\,\theta_{23})\right)\sin(\Delta)\right.\nonumber \\
 &  & \left.-\left(\widehat{A}_{e}+\cos(2\,\theta_{23})\right)\,\sin(\Delta-2\,\widehat{A}_{e}\,\Delta)\right]-\left(-1+\widehat{A}_{e}\right)\,\sin(\delta_{13}+\chi_{e\tau})\,\nonumber \\
 &  & \left[\,\sin(2\,\Delta)-\sin(2\,(1-\widehat{A}_{e})\,\Delta)-\sin(2\,\widehat{A}_{e}\,\Delta)\right]\biggr]\biggr\}\,.\nonumber \end{eqnarray}
 For the mass correction terms, the first one $C_{\mu\tau}^{0}$ is
given by Eq.(\ref{E232}), and the two other universal terms are the
following: \begin{equation}
C_{\mu\tau}^{\alpha}=C^{\alpha}\,,\,\,\,\, C_{\mu\tau}^{s}=C^{s}\,.\label{A10}\end{equation}
 Also for the $\nu_{\mu}\rightarrow\nu_{\mu}$ transition, all terms
contribute. The modulus of the first interaction correction term is
given by Eq.(\ref{E232}), whereas the two remaining terms are as
follows: \begin{eqnarray}
B_{\mu\mu}^{\alpha} & = & \frac{2}{\left(-1+\widehat{A}_{e}\right)\,\widehat{A}_{e}}\,\biggl\{\varepsilon_{e\mu}\,\cos(\chi_{e\mu})\,\cos(\theta_{23})\,\sin(2\,\theta_{12})\,\left[\left(-1+\widehat{A}_{e}\right)\,\right.\label{A11}\\
 &  & \left.\cos(2\,\widehat{A}_{e}\,\Delta)\,\cos^{4}(\theta_{23})+\widehat{A}_{e}\,\sin^{2}(\theta_{23})\,\left(-1+\cos(2\,(-1+\widehat{A}_{e})\,\Delta)\,\sin^{2}(\theta_{23})\right.\right.\nonumber \\
 &  & \left.\left.-2\,(-1+\widehat{A}_{e})\,\sin^{2}(\Delta)\,\sin^{2}(\theta_{23})\right)+\cos^{2}(\theta_{23})\,\left(1-\widehat{A}_{e}\right.\right.\nonumber \\
 &  & \left.\left.+(-1+\widehat{A}_{e})\,\cos(2\,(-1+\widehat{A}_{e})\,\Delta)\,\sin^{2}(\theta_{23})+\widehat{A}_{e}\,\cos(2\,\widehat{A}_{e}\,\Delta)\,\sin^{2}(\theta_{23})\right.\right.\nonumber \\
 &  & \left.\left.-2\,\sin^{2}(\Delta)\,\sin^{2}(\theta_{23})+2\,\widehat{A}_{e}\,\sin^{2}(\Delta)\,\sin^{2}(\theta_{23})+2\,\widehat{A}_{e}^{2}\,\sin^{2}(\Delta)\,\sin^{2}(\theta_{23})\right)\right]\,\nonumber \\
 &  & +\frac{1}{2}\varepsilon_{e\tau}\,\cos(\chi_{e\tau})\,\cos(\theta_{23})\,\sin(2\,\theta_{12})\,\nonumber \\
 &  & \left[2\,\cos(2\,\widehat{A}_{e}\,\Delta)\,\cos^{3}(\theta_{23})\,\sin(\theta_{23})+2\,\cos(2\,(-1+\widehat{A}_{e})\,\Delta)\,\cos(\theta_{23})\,\sin^{3}(\theta_{23})\right.\nonumber \\
 &  & \left.+4\,\cos(\theta_{23})\,\sin^{2}(\Delta)\,\sin^{3}(\theta_{23})-\sin(2\,\theta_{23})+\widehat{A}_{e}^{2}\,\sin^{2}(\Delta)\,\sin(4\,\theta_{23})\right]\,\nonumber \\
 &  & +\left(-1+\widehat{A}_{e}\right)\,\widehat{A}_{e}^{2}\,\cos^{2}(\theta_{12})\,\left(2\,\Delta\,\cos(\Delta)-\sin(\Delta)\right)\,\sin(\Delta)\,\sin(4\,\theta_{23})\,\nonumber \\
 &  & \left[\sin(2\,\theta_{23})\,(\varepsilon_{\mu\mu}-\varepsilon_{\tau\tau})+2\,\cos(2\,\theta_{23})\,\cos(\chi_{\mu\tau})\,\varepsilon_{\mu\tau}\right]\biggr\}\,,\nonumber \end{eqnarray}
 \begin{eqnarray}
B_{\mu\mu}^{s} & = & \frac{2}{\left(-1+\widehat{A}_{e}\right)^{2}}\,\biggl\{\varepsilon_{e\mu}\,\cos(\delta_{13}+\chi_{e\mu})\,\biggl[\,\frac{\sin(\theta_{23})}{2}\left(-1+2\widehat{A}_{e}-\cos(2\,\theta_{23})\right)\,\,\label{A12}\\
 &  & \left.\left(-1+\cos(2\,\widehat{A}_{e}\,\Delta)\,\cos^{2}(\theta_{23})+\cos(2\,(-1+\widehat{A}_{e})\,\Delta)\,\sin^{2}(\theta_{23})\right)\right.\nonumber \\
 &  & +\cos(\theta_{23})\,\sin^{2}(\Delta)\,\left((-1+\widehat{A}_{e})\,\widehat{A}_{e}\,\cos^{2}(\theta_{23})-(1+(-3+\widehat{A}_{e})\,\widehat{A}_{e})\,\sin^{2}(\theta_{23})\right)\,\,\,\nonumber \\
 &  & \sin(2\,\theta_{23})\biggr]+\varepsilon_{e\tau}\,\cos(\delta_{13}+\chi_{e\tau})\,\sin(\theta_{23})\,\nonumber \\
 &  & \left[-2\,\cos^{3}(\theta_{23})\,\sin^{2}(\widehat{A}_{e}\,\Delta)\,\sin(\theta_{23})-2\,\cos(\theta_{23})\,\sin^{2}(\Delta-\widehat{A}_{e}\,\Delta)\,\sin^{3}(\theta_{23})\right.\nonumber \\
 &  & \left.+\sin^{2}(\Delta)\,\left((2-\widehat{A}_{e})\,\widehat{A}_{e}\,\cos(2\,\theta_{23})+\sin^{2}(\theta_{23})\right)\,\sin(2\,\theta_{23})\right]\biggr\}\,.\nonumber \end{eqnarray}
 The mass term for this channel has the opposite sign, in comparison
to the previously discussed transition $\nu_{\mu}\rightarrow\nu_{\tau}$:
\begin{equation}
\delta P_{\mu\rightarrow\mu}^{mass}\,=-\delta P_{\mu\rightarrow\tau}^{mass}\,.\label{A13}\end{equation}
 Finally, the $\nu SM$ probabilities are as follows: \begin{eqnarray}
P_{\mu\rightarrow e}^{SM} & = & \frac{\alpha^{2}}{\widehat{A}_{e}^{2}}\left(\,\cos^{2}(\theta_{23})\,\sin^{2}(\widehat{A}_{e}\,\Delta)\,\sin^{2}(2\,\theta_{12})\right)\label{A14}\\
 &  & +\frac{\sin^{2}(2\,\theta_{13})}{\left(-1+\widehat{A}_{e}\right)^{2}}\left(\sin^{2}(\Delta-\widehat{A}_{e}\,\Delta)\,\sin^{2}(\theta_{23})\right)\nonumber \\
 &  & -\frac{\alpha\,\sin(2\,\theta_{13})}{\left(-1+\widehat{A}_{e}\right)\,\widehat{A}_{e}}\,\biggl\{\left[\cos(\delta_{13})\,\cos(\Delta)+\sin(\delta_{13})\,\sin(\Delta)\right]\,\nonumber \\
 &  & \sin(\widehat{A}_{e}\,\Delta)\,\sin(\Delta-\widehat{A}_{e}\,\Delta)\sin(2\,\theta_{12})\,\sin(2\,\theta_{23})\biggr\}\,,\nonumber \end{eqnarray}
 \begin{eqnarray}
P_{\mu\rightarrow\tau}^{SM} & = & \sin^{2}(\Delta)\,\sin^{2}(2\,\theta_{23})-\alpha\,\Delta\,\cos^{2}(\theta_{12})\,\sin(2\,\Delta)\,\sin^{2}(2\,\theta_{23})\label{A15}\\
 &  & +\frac{\alpha^{2}}{4\,\widehat{A}_{e}^{2}}\,\biggl\{\sin(\Delta)\,\sin(\Delta-2\,\widehat{A}_{e}\,\Delta)\,\sin^{2}(2\,\theta_{12})\,\sin^{2}(2\,\theta_{23})\biggr\}\nonumber \\
 &  & -\frac{\sin^{2}(2\,\theta_{13})}{4\,\left(-1+\widehat{A}_{e}\right)^{2}}\,\biggl\{\sin(\Delta)\,\left[2\,\left(\widehat{A}_{e}-1\right)\,\widehat{A}_{e}\,\Delta\,\cos(\Delta)\right.\nonumber \\
 &  & \left.+\sin(\Delta)+\sin(\Delta-2\,\widehat{A}_{e}\,\Delta)\right]\,\sin^{2}(2\,\theta_{23})\biggr\}\,\nonumber \\
 &  & +\frac{\alpha\,\sin(2\,\theta_{13})}{2\,\left(-1+\widehat{A}_{e}\right)\,\widehat{A}_{e}}\,\biggl\{-\cos(\delta_{13})\,\biggl[\cos(2\,\theta_{23})\,\sin(\Delta)\nonumber \\
 &  & \left(\left(-1+2\,\widehat{A}_{e}^{2}\right)\,\sin(\Delta)+\sin(\Delta-2\,\widehat{A}_{e}\,\Delta)\right)\,\sin(2\,\theta_{12})\,\sin(2\,\theta_{23})\,\biggr]\nonumber \\
 &  & -2\,\sin(\delta_{13})\,\biggl[\sin(\Delta)\,\sin(\widehat{A}_{e}\,\Delta)\,\sin(\Delta-\widehat{A}_{e}\,\Delta)\,\sin(2\,\theta_{12})\,\sin(2\,\theta_{23})\biggr]\biggr\}\,,\nonumber \end{eqnarray}
 \begin{eqnarray}
P_{\mu\rightarrow\mu}^{SM} & = & 1-\sin^{2}(\Delta)\,\sin^{2}(2\,\theta_{23})+\alpha\,\Delta\,\cos^{2}(\theta_{12})\,\sin(2\,\Delta)\,\sin^{2}(2\,\theta_{23})\label{A16}\\
 &  & +\frac{\sin^{2}(2\,\theta_{13})}{4\,\left(-1+\widehat{A}_{e}\right)^{2}}\,\biggl\{-4\,\sin^{2}(\Delta-\widehat{A}_{e}\,\Delta)\,\sin^{4}(\theta_{23})\nonumber \\
 &  & +\left(\sin^{2}(\Delta)+\left(-1+\widehat{A}_{e}\right)\,\widehat{A}_{e}\,\Delta\,\sin(2\,\Delta)-\sin^{2}(\widehat{A}_{e}\,\Delta)\right)\,\sin^{2}(2\,\theta_{23})\biggr\}\nonumber \\
 &  & -\frac{\alpha^{2}}{4\,\widehat{A}_{e}^{2}}\,\sin^{2}(2\,\theta_{12})\,\biggl\{4\,\cos^{4}(\theta_{23})\,\sin^{2}(\widehat{A}_{e}\,\Delta)+\sin^{2}(\Delta-\widehat{A}_{e}\,\Delta)\,\sin^{2}(2\,\theta_{23})\biggr\}\nonumber \\
 &  & +\frac{\alpha\,\sin(2\,\theta_{13})}{\left(-1+\widehat{A}_{e}\right)\,\widehat{A}_{e}}\,\cos(\delta_{13})\,\sin(2\,\theta_{12})\,\biggl\{\sin^{2}(\Delta)\,\left(\widehat{A}_{e}^{2}\,\cos(2\,\theta_{23})+\sin^{2}(\theta_{23})\right)\,\sin(2\,\theta_{23})\nonumber \\
 &  & -\sin(\theta_{23})\,\left(2\,\cos^{3}(\theta_{23})\,\sin^{2}(\widehat{A}_{e}\,\Delta)+\sin^{2}(\Delta-\widehat{A}_{e}\,\Delta)\,\sin(\theta_{23})\,\sin(2\,\theta_{23})\right)\biggr\}\,.\nonumber \end{eqnarray}

The formulas for the time reversed channels, $\nu_{e}\rightarrow\nu_{\mu}$
and $\nu_{\tau}\rightarrow\nu_{\mu}$, are obtained from the above
formulas after the replacements $\Delta\rightarrow-\Delta$ and $\widetilde{\Delta}_{ik}\rightarrow-\widetilde{\Delta}_{ik}$.
Finally, if no scalar nor pseudoscalar neutrino interactions are considered
(like in the model which we present here, see also \cite{heavy}),
for antineutrinos the interaction Hamiltonian $H$ is replaced by
$-H^{*}$, thus the transitions formulas for channels $\nu_{\bar{\mu}}\rightarrow\nu_{\bar{e}}$
, $\nu_{\bar{\mu}}\rightarrow\nu_{\bar{\tau}}$ and $\nu_{\bar{\mu}}\rightarrow\nu_{\bar{\mu}}$
are easily obtained after the replacements $\delta_{13}\rightarrow-\delta_{13}$,
$\widehat{A}_{e}\rightarrow-\widehat{A}_{e}$, and $\delta\lambda_{\alpha\beta}\rightarrow\delta\lambda_{\alpha\beta}^{*}$
(hence also $\chi_{\alpha\beta}\rightarrow-\chi_{\alpha\beta}$).
This completes the set of $\nu SM$ transition probability formulas
and the NP corrections to them, for all experimentally available channels.

\end{widetext}

\end{document}